\documentclass[reprint,
amsmath,
amssymb,
aps
]{revtex4-1}

\usepackage{graphicx}
\usepackage{bm}
\usepackage{hyperref}
\usepackage{amsmath}
\usepackage{braket}
\usepackage{bbding}
\usepackage{xcolor}
\usepackage{comment}
\newcommand{\dd}[1]{\text{d}#1}
\newcommand{\dr}{\text{d}^{3}\bm{r}}
\begin{document}

\title{Degenerate Squeezing in Waveguides: A Unified Theoretical Approach}

\author{L. G. Helt}
\affiliation{Xanadu, Toronto, Ontario, Canada M5G 2C8}
\author{N. Quesada}
\affiliation{Xanadu, Toronto, Ontario, Canada M5G 2C8}

\date{\today}

\begin{abstract}
We consider pulsed-pump spontaneous parametric downconversion (SPDC) as well as pulsed single- and dual-pump spontaneous four-wave mixing processes in waveguides within a unified Hamiltonian theoretical framework. Working with linear operator equations in $k$-space, our approach allows inclusion of linear losses, self- and cross-phase modulation, and dispersion to any order. We describe state evolution in terms of second-order moments, for which we develop explicit expressions. We use our approach to calculate the joint spectral amplitude of degenerate squeezing using SPDC analytically in the perturbative limit, benchmark our theory against well-known results in the limit of negligible group velocity dispersion, and study the suitability of recently proposed sources for quantum sampling experiments. 
\end{abstract}

\pacs{Valid PACS appear here}

\maketitle

\section{Introduction}
First observed in the 1980s~\cite{slusher1985observation,shelby1986broadband,wu1986generation,machida1987observation}, squeezed states of light have become more ubiquitous over time~\cite{andersen201630}. While part of this is due to fundamental interest, there are many practical applications of squeezed light as well, including computation~\cite{yoshikawa2018optical}, communication~\cite{laudenbach2018continuous}, and metrology~\cite{pirandola2018advances}. In recent years, Gaussian Boson Sampling (GBS)~\cite{hamilton2017gaussian,kruse2018a} has emerged as one of the most actively researched applications~\cite{arrazola2018quantum, arrazola2018using, gupt2018classical, quesada2018gaussian, zhong2019experimental, banchi2019molecular, jahangiri2019point,schuld2019quantum}.

GBS involves a number of Gaussian states (i.e. those with a Gaussian Wigner function) input to an interferometer and subsequently detected at the output of the interferometer with photon number resolving (PNR) detectors. For non-classical input Gaussian states, sampling from the output distribution of photon number events at the detectors is conjectured to be a \#P hard problem and thus intractable by classical computers for a large enough number of input states~\cite{neville2017classical, clifford2018classical,quesada2019classical,wu2019speedup}. This is interesting from a theoretical point of view, for it stands to provide evidence against the extended Church-Turing thesis~\cite{parberry1986parallel}. It is also experimentally interesting, for, in contrast to heralded single photon states, non-classical Gaussian states such as squeezed states can be created deterministically and so medium-scale implementations of GBS could be possible in the near future.

However, it should be noted that not every squeezed state of light is useful for GBS. In particular, the \#P hardness depends on collecting samples in the photon number basis from states in a single known mode, and PNR detectors are not fast enough to provide spectral mode selectivity~\cite{du2015quantum}. To get around this, a well defined spectral mode can be arranged by employing a pulsed pump in an appropriate nonlinear medium~\cite{vernon2018scalable}. Common nonlinear media include second- and third-order nonlinear optical waveguides, capable of producing squeezed states via spontaneous parametric downconversion (SPDC) or spontaneous four-wave mixing (SFWM), respectively. We note that this situation is different from both the perturbative photon pair generation regime~\cite{mosley2008heralded}, as well as that of a continuous wave pump followed by homodyne detection~\cite{yonezawa2004demonstration}. Additionally, degenerate, rather than non-degenerate or so called twin-beam squeezing, is considered in studies of the hardness of GBS~\cite{hamilton2017gaussian,kruse2018a} and its applications~\cite{arrazola2018quantum, arrazola2018using, gupt2018classical, quesada2018gaussian, zhong2019experimental, banchi2019molecular, jahangiri2019point,schuld2019quantum}.

In this work we therefore develop theoretical tools to calculate degenerate squeezing generated by pulsed pumps in waveguides. In contrast to some previous approaches, we focus on pulsed, rather than CW~\cite{shapiro1995raman, voss2006raman, mckinstrie2011field}, pumps, avoid phase-space methods~\cite{drummond2001quantum}, and work with operator coupled mode equations driven by first-order temporal, rather than spatial, \cite{kolobov1999spatial,mckinstrie2007phase, lin2007photon, caves1987quantum, brainis2009four, bell2015effects, koefoed2019complete,lai1995general,christ2013theory}, derivatives.  In particular, we note that coupled mode equations in which a series expansion in $\tfrac{\partial}{\partial t}$ is used to accommodate material and modal dispersion, often used in the nonlinear optics community for classical fields, are not derived from a canonical formalism and thus it is not obvious that they are valid in the quantum regime~\footnote{Indeed, as noted by Dirac in Sec.~67 of Ref.~\cite{dirac1948principles}, one can deduce ``from quite general arguments that the [quantum-mechanical] wave equation must be linear in the operator $\tfrac{\partial}{\partial t}$''.}%{diracnote, dirac1948principles}.

To provide a treatment consistent with Maxwell's equations and quantum mechanics we start from an established Hamiltonian formalism in Section~\ref{sec:formalism}, where we use the Heisenberg equations of motion to obtain operator coupled mode equations. In subsections contained therein, we massage these equations into a form useful for numerical evaluation, introduce a simple method to include loss, and work out expressions for the phase sensitive and phase insensitive moments, generated state, and the state's associated joint spectral amplitude (JSA). Having established these tools, in Section~\ref{sec:examples} we consider three example calculations: SPDC in the low-gain regime, homodyne detection of single-pump SFWM as a function of gain, and dual-pump SFWM in the high-gain regime. The first two examples serve as reasonable checks of our formalism, for they have been studied previously~\cite{mosley2008heralded, shirasaki1990squeezing}, while the final example is most relevant for GBS. We conclude in section~\ref{sec:conclusions}.

\section{Formalism}
\label{sec:formalism}
We follow a canonical formalism \cite{yang2008spontaneous,sipe2009photons,liscidini2012asymptotic,quesada2017why,drummond2014quantum,quesada2019efficient}, which accounts for material and modal dispersion and yields Heisenberg equations of motion for the operator fields $\bm{D}\left(\bm{r}\right)$ and $\bm{B}\left(\bm{r}\right)$ that reproduce the dynamical Maxwell equations. This approach allows results of calculations to be absolute, rather than relative to an unknown constant, and facilitates simple comparisons with both experimental results and classical calculations. Waveguide propagation is assumed to be in the positive $z$ direction. We begin with linear (L) and nonlinear (NL) Hamiltonians
\begin{equation}
H=H_{L}+H_{NL}^{\left(2\right)}+H_{NL}^{\left(3\right)}+\cdots,
\end{equation}
where
\begin{equation}
H_{L}=\sum_{J}\int\dd{k}\,\hbar\omega_{Jk}a_{Jk}^{\dagger}a_{Jk},
\label{eq:HLa}
\end{equation}
\begin{equation}
H_{NL}^{\left(2\right)}=-\frac{1}{3\varepsilon_{0}}\int\dr\,\Gamma_{jlm}^{\left(2\right)}\left(\bm{r}\right)D^{j}\left(\bm{r}\right)D^{l}\left(\bm{r}\right)D^{m}\left(\bm{r}\right),
\end{equation}
and
\begin{align}
H_{NL}^{\left(3\right)}&=-\frac{1}{4\varepsilon_{0}}\int\dr\,\Gamma_{jlmn}^{\left(3\right)}\left(\bm{r}\right)D^{j}\left(\bm{r}\right)D^{l}\left(\bm{r}\right)\nonumber\\
&\qquad\times D^{m}\left(\bm{r}\right)D^{n}\left(\bm{r}\right),
\end{align}
with
\begin{equation}
\left[a_{Jk},a_{J^{\prime}k^{\prime}}^{\dagger}\right]=\delta_{JJ^{\prime}}\delta\left(k-k^{\prime}\right).
\end{equation}
Note that the $\Gamma^{\left(n\right)}\left(\bm{r}\right)$ are $n$th-rank tensors linearly related to the more common susceptibility tensors $\chi^{\left(n\right)}\left(\bm{r}\right)$ (see Appendix \ref{app:CMEs} for details).

We associate relevant waveguide modes with center wavevectors $k_{J}$ and frequencies $\bar{\omega}_{J}=\omega_{Jk_{J}}$. Taylor expanding
\begin{equation}\label{eq:disprel}
\omega_{Jk}=\bar{\omega}_J+v_{J}\left(k-k_{J}\right)+\frac{v_{J}^{\prime}}{2}\left(k-k_{J}\right)^{2}+\cdots,
\end{equation}
where
\begin{equation}
v_{J}=\left.\frac{\dd{\omega_{Jk}}}{\dd{k}}\right\vert_{k=k_{J}},\quad v_{J}^\prime=\left.\frac{\text{d}^{2}\omega_{Jk}}{\dd{k}^{2}}\right\vert_{k=k_{J}},
\end{equation}
and introducing Fourier transformed operators centered at $k_{J}$ such that they are slowly-varying in space
\begin{equation}
\psi_{J}\left(z\right)=\int\dd{k}\frac{e^{i\left(k-k_{J}\right)z}}{\sqrt{2\pi}}a_{Jk},
\label{eq:psiz}
\end{equation}
allows us to rewrite Eq.~\eqref{eq:HLa} as
\begin{align}
H_{L}=&\sum_{J}\int\dd{z}\,\hbar\left[\vphantom{\frac{\partial\psi_{J}^{\dagger}\left(z\right)}{\partial z}}\bar{\omega}_{J}\psi_{J}^{\dagger}\left(z\right)\psi_{J}\left(z\right)\right.\nonumber\\
&\quad+\frac{i}{2}v_{J}\left[\frac{\partial\psi_{J}^{\dagger}\left(z\right)}{\partial z}\psi_{J}\left(z\right)-\psi_{J}^{\dagger}\left(z\right)\frac{\partial\psi_{J}\left(z\right)}{\partial z}\right]\nonumber\\
&\quad\left.+\frac{v_{J}^{\prime}}{2}\frac{\partial\psi_{J}^{\dagger}\left(z\right)}{\partial z}\frac{\partial\psi_{J}\left(z\right)}{\partial z}+\cdots\right].
\label{eq:HLpsi}
\end{align}
Note that these new operators satisfy simple commutation relations 
\begin{equation}
\left[\psi_{J}\left(z\right),\psi_{J^{\prime}}^{\dagger}\left(z^{\prime}\right)\right]=\delta_{JJ^{\prime}}\delta\left(z-z^{\prime}\right),
\end{equation}
to excellent approximation (as they are exact only assuming that the modes exist for all $k$) for the $\psi_{J}\left(z\right)$ of interest. However the real power of the $\psi_{J}\left(z\right)$ is in the approximations they allow for $\bm{D}\left(\bm{r}\right)$ and thus the nonlinear Hamiltonians.

In general, the displacement field is expanded in terms of waveguide modes $J$ in the transverse plane $\bm{d}_{Jk}\left(x,y\right)$ as
\begin{equation}
\bm{D}\left(\bm{r}\right)=\sum_{J}\int\text{d}k\,\sqrt{\frac{\hbar\omega_{Jk}}{2}}\bm{d}_{Jk}\left(x,y\right)\frac{e^{ikz}}{\sqrt{2\pi}}a_{Jk}+\text{H.c.}
\label{eq:fullD}
\end{equation}
Using Eq.~\eqref{eq:psiz}, however, we may form a useful approximate expression for Eq.~\eqref{eq:fullD} as
\begin{equation}
\bm{D}\left(\bm{r}\right)=\sum_{J}e^{ik_{J}z}\sqrt{\frac{\hbar\bar{\omega}_{J}}{2}}\bm{d}_{Jk_{J}}\left(x,y\right)\psi_{J}\left(z\right)+\text{H.c.}+\cdots.
\label{eq:simpleD}
\end{equation}
Because the nonlinearity is weak, while we include dispersion in the linear Hamiltonian, we comfortably neglect dispersion in the nonlinear Hamiltonians (aside from in the normalization of the $\bm{d}_{Jk_{J}}\left(x,y\right)$~\cite{yang2008spontaneous,sipe2009photons}) and use only the first term of Eq.~\eqref{eq:simpleD} in their construction. This allows simple identification of phase matched terms in nonlinear optical processes of interest.

With Hamiltonians defined, we are now in a position to study system dynamics. One approach is to solve the Schr\"odinger equation for the time evolution operator
\begin{align}\label{eq:defU}
i \hbar \frac{\dd}{\dd t} \mathcal{U}(t,t_0) = \left(  H_L +H_{NL} \right)  \mathcal{U}(t,t_0),
\end{align}
satisfying $\mathcal{U}(t,t) = \mathbb{I}$, and subsequently use this operator to transform kets in the usual way $\ket{\Psi(t)} = \mathcal{U}(t,t_0) \ket{\Psi(t_0)}$. However, we can also transform Schr\"odinger operators to time dependent Heisenberg operators
\begin{align}\label{eq:Heisenberg}
\psi(z,t) &= \mathcal{U}^\dagger(t,t_0) \psi(z) \mathcal{U}(t,t_0),\\
a(k,t) &= \mathcal{U}^\dagger(t,t_0) a(k) \mathcal{U}(t,t_0),
\end{align}
and solve their equations of motion instead. In the remainder of this Section we derive equations for these Heisenberg operators and show how to find time evolved kets using these solutions in the important case where the initial state $\ket{\Psi(t_0)} = \ket{\text{vac}}$.

\subsection{Lossless coupled-mode equations}
Here we use the expressions from the start of the Section to develop nonlinear Hamiltonians for degenerate squeezing processes, and subsequently use these Hamiltonians to derive operator coupled-mode mode equations in a common form. Details can be found in Appendix \ref{app:CMEs}.

For SPDC, we take $J\rightarrow\left\{F,SH\right\}$  for ``fundamental'' and ``second harmonic'', consider only the second-order nonlinearity $H_{NL}\rightarrow H_{NL}^{\left(2\right)}$, and, using only the first term of Eq.~\eqref{eq:simpleD}, write
\begin{equation}
H_{NL}=-\frac{\hbar}{2}\int\dd{z}\,\zeta^{\left(2\right)}\left(z\right)\psi_{F}^{\dagger}\left(z\right)\psi_{F}^{\dagger}\left(z\right)\psi_{SH}\left(z\right)+\text{H.c.},
\label{eq:HSPDC}
\end{equation}
where $\zeta^{\left(2\right)}\left(z\right)$ is a nonlinear coupling parameter defined in Appendix \ref{app:CMEs}, and we have assumed that the only phase (energy) matched process occurs for $2k_{F}=k_{SH}$ ($2\bar{\omega}_{F} = \bar{\omega}_{SH}$). Note that the $z$-dependence of the nonlinear coupling parameter simply defines the nonlinear region of the waveguide. Coupled mode equations then follow immediately from the Heisenberg equations of motion $\partial_{t}O=i\left[H_{L}+H_{NL},O\right]/\hbar$. To avoid propagating oscillations at optical frequencies in the dynamics, we calculate the Heisenberg equations of motion for the slowly oscillating operators
\begin{align}
\bar{\psi}_J(z,t) = e^{i \bar{\omega}_J t} \psi_J(z,t),
\end{align}
obtaining
\begin{equation}
\left(\frac{\partial}{\partial t}+v_{SH}\frac{\partial}{\partial z}-i\frac{v_{SH}^{\prime}}{2}\frac{\partial^{2}}{\partial z^{2}}\right)\left\langle\bar{\psi}_{SH}\left(z,t\right)\right\rangle=0,
\label{eq:psiSH}
\end{equation}
and
\begin{align}
&\left(\frac{\partial}{\partial t}+v_{F}\frac{\partial}{\partial z}-i\frac{v_{F}^{\prime}}{2}\frac{\partial^{2}}{\partial z^{2}}\right)\bar{\psi}_{F}\left(z,t\right)\nonumber\\
&\qquad=i\zeta^{\left(2\right)}\left(z\right)\left\langle\bar{\psi}_{SH}\left(z,t\right)\right\rangle\bar{\psi}_{F}^{\dagger}\left(z,t\right),
\label{eq:psiF}
\end{align}
where we have considered the driving field at the second harmonic to be a large classical field and thus replaced all instances of $\bar{\psi}_{SH}\left(z\right)$ by its mean field $\left\langle \bar{\psi}_{SH}\left(z\right)\right\rangle$ and neglected any back-action on it from $\bar{\psi}_{F}\left(z\right)$. For clarity of presentation, we do not include higher-order dispersion or the effects of self- or cross-phase modulation (SPM and XPM, respectively) here. However, we note that our formalism admits consideration of higher-order dispersion simply by including more terms in Eq.~\eqref{eq:HLpsi}, as well as consideration of SPM and XPM by allowing $H_{NL}=H_{NL}^{\left(2\right)}+H_{NL}^{\left(3\right)}$.

For single-pump SFWM, we have only a single mode $J\rightarrow P$ for ``pump'', consider only the third-order nonlinearity $H_{NL}\rightarrow H_{NL}^{\left(3\right)}$, and, using only the first term of Eq.~\eqref{eq:simpleD}, write
\begin{equation}
H_{NL}=-\frac{\hbar}{2}\int\dd{z}\,\zeta_{PPPP}^{\left(3\right)}\left(z\right)\psi_{P}^{\dagger}\left(z\right)\psi_{P}^{\dagger}\left(z\right)\psi_{P}\left(z\right)\psi_{P}\left(z\right),
\label{eq:HSPSFWM}
\end{equation}
where $\zeta_{J_{1}J_{2}J_{3}J_{4}}^{\left(3\right)}\left(z\right)$	is a nonlinear coupling parameter defined in Appendix \ref{app:CMEs}. From the Heisenberg equations of motion we find
\begin{align}
&\left(\frac{\partial}{\partial t}+v_{P}\frac{\partial}{\partial z}-i\frac{v_{P}^{\prime}}{2}\frac{\partial^{2}}{\partial z^{2}}\right)\left\langle \bar{\psi}_{P}\left(z,t\right)\right\rangle\nonumber\\ &\qquad=i\zeta_{PPPP}^{\left(3\right)}\left(z\right)\left|\left\langle \bar{\psi}_{P}\left(z,t\right)\right\rangle \right|^{2}\left\langle \bar{\psi}_{P}\left(z,t\right)\right\rangle,
\label{eq:psiP}
\end{align}
and
\begin{align}
&\left(\frac{\partial}{\partial t}+v_{P}\frac{\partial}{\partial z}-i\frac{v_{P}^{\prime}}{2}\frac{\partial^{2}}{\partial z^{2}}\right)\delta\bar{\psi}_{P}\left(z,t\right)\nonumber\\
&\qquad=i\zeta_{PPPP}^{\left(3\right)}\left(z\right)\left\langle \bar{\psi}_{P}\left(z,t\right)\right\rangle ^{2}\delta\bar{\psi}_{P}^{\dagger}\left(z,t\right)\nonumber\\
&\qquad\quad+2i\zeta_{PPPP}^{\left(3\right)}\left(z\right)\left|\left\langle \bar{\psi}_{P}\left(z,t\right)\right\rangle \right|^{2}\delta\bar{\psi}_{P}\left(z,t\right),
\label{eq:dpsiP}
\end{align}
where we have considered $\bar{\psi}_{P}\left(z,t\right)$ as its mean field plus fluctuations $\bar{\psi}_{P}\left(z,t\right)\rightarrow\left\langle \bar{\psi}_{P}\left(z,t\right)\right\rangle +\delta\bar{\psi}_{P}\left(z,t\right)$, and neglected terms quadratic in the fluctuations or higher.

For dual-pump SFWM, we take $J\rightarrow\left\{P_{1},P_{2},S\right\}$  for ``pump 1'', ``pump 2'' and ``signal'', consider only the third-order nonlinearity $H_{NL}\rightarrow H_{NL}^{\left(3\right)}$, and, using only the first term of Eq.~\eqref{eq:simpleD}, write
\begin{align}\
&H_{NL} \label{eq:HDPSFWM} \\	
&=-\hbar\int\dd{z}\,\zeta_{SSP_{1}P_{2}}^{\left(3\right)}\left(z\right)\psi_{S}^{\dagger}\left(z\right)\psi_{S}^{\dagger}\left(z\right)\psi_{P_{1}}\left(z\right)\psi_{P_{2}}\left(z\right)+\text{H.c.}\nonumber\\
&\quad-\frac{\hbar}{2}\sum_{J}\int\dd{z}\,\zeta_{JJJJ}^{\left(3\right)}\left(z\right)\psi_{J}^{\dagger}\left(z\right)\psi_{J}^{\dagger}\left(z\right)\psi_{J}\left(z\right)\psi_{J}\left(z\right)\nonumber\\
&\quad-2\hbar\sum_{J\neq S}\int\dd{z}\,\zeta_{SJSJ}^{\left(3\right)}\left(z\right)\psi_{S}^{\dagger}\left(z\right)\psi_{J}^{\dagger}\left(z\right)\psi_{S}\left(z\right)\psi_{J}\left(z\right)\nonumber\\
&\quad-2\hbar\int\dd{z}\,\zeta_{P_{1}P_{2}P_{1}P_{2}}^{\left(3\right)}\left(z\right)\psi_{P_{1}}^{\dagger}\left(z\right)\psi_{P_{2}}^{\dagger}\left(z\right)\psi_{P_{1}}\left(z\right)\psi_{P_{2}}\left(z\right), \nonumber
\end{align}
where $\zeta_{J_{1}J_{2}J_{3}J_{4}}^{\left(3\right)}\left(z\right)$	is a nonlinear coupling parameter defined in Appendix \ref{app:CMEs}, and we have assumed $2k_{S}=k_{P_{1}}+k_{P_{2}}$ and $2\bar{\omega}_{S}=\bar{\omega}_{P_{1}}+\bar{\omega}_{P_{2}}$.
Coupled mode equations then follow from the Heisenberg equations of motion as
\begin{align}
&\left(\frac{\partial}{\partial t}+v_{P_{1}}\frac{\partial}{\partial z}-i\frac{v_{P_{1}}^{\prime}}{2}\frac{\partial^{2}}{\partial z^{2}}\right)\left\langle \bar{\psi}_{P_{1}}\left(z,t\right)\right\rangle\nonumber\\ &\qquad=i\zeta_{P_{1}P_{1}P_{1}P_{1}}^{\left(3\right)}\left(z\right)\left|\left\langle \bar{\psi}_{P_{1}}\left(z,t\right)\right\rangle \right|^{2}\left\langle \bar{\psi}_{P_{1}}\left(z,t\right)\right\rangle\nonumber\\ &\qquad\quad+2i\zeta_{P_{1}P_{2}P_{1}P_{2}}^{\left(3\right)}\left(z\right)\left|\left\langle \bar{\psi}_{P_{2}}\left(z,t\right)\right\rangle \right|^{2}\left\langle \bar{\psi}_{P_{1}}\left(z,t\right)\right\rangle,
\label{eq:psiP1}
\end{align}
\begin{align}
&\left(\frac{\partial}{\partial t}+v_{P_{2}}\frac{\partial}{\partial z}-i\frac{v_{P_{2}}^{\prime}}{2}\frac{\partial^{2}}{\partial z^{2}}\right)\left\langle \bar{\psi}_{P_{2}}\left(z,t\right)\right\rangle\nonumber\\ &\qquad=i\zeta_{P_{2}P_{2}P_{2}P_{2}}^{\left(3\right)}\left(z\right)\left|\left\langle \bar{\psi}_{P_{2}}\left(z,t\right)\right\rangle \right|^{2}\left\langle \bar{\psi}_{P_{2}}\left(z,t\right)\right\rangle\nonumber\\
&\qquad\quad+2i\zeta_{P_{1}P_{2}P_{1}P_{2}}^{\left(3\right)}\left(z\right)\left|\left\langle \bar{\psi}_{P_{1}}\left(z,t\right)\right\rangle \right|^{2}\left\langle \bar{\psi}_{P_{2}}\left(z,t\right)\right\rangle,
\label{eq:psiP2}
\end{align}
and
\begin{align}
&\left(\frac{\partial}{\partial t}+v_{S}\frac{\partial}{\partial z}-i\frac{v_{S}^{\prime}}{2}\frac{\partial^{2}}{\partial z^{2}}\right)\bar{\psi}_{S}\left(z,t\right)\nonumber\\
&\qquad=2i\zeta_{SSP_{1}P_{2}}^{\left(3\right)}\left(z\right)\left\langle \bar{\psi}_{P_{1}}\left(z,t\right)\right\rangle \left\langle \bar{\psi}_{P_{2}}\left(z,t\right)\right\rangle \bar{\psi}_{S}^{\dagger}\left(z,t\right)\nonumber\\
&\qquad\quad+2i\zeta_{SP_{2}SP_{2}}^{\left(3\right)}\left(z\right)\left|\left\langle \bar{\psi}_{P_{2}}\left(z,t\right)\right\rangle \right|^{2}\bar{\psi}_{S}\left(z,t\right)\nonumber\\
&\qquad\quad+2i\zeta_{SP_{1}SP_{1}}^{\left(3\right)}\left(z\right)\left|\left\langle \bar{\psi}_{P_{1}}\left(z,t\right)\right\rangle \right|^{2}\bar{\psi}_{S}\left(z,t\right),
\label{eq:psiS}
\end{align}
where we have considered the driving fields labeled by $P_{1}$ and $P_{2}$ to be a large classical fields and thus replaced all instances of $\bar{\psi}_{P_{1}}\left(z,t\right)$ and $\bar{\psi}_{P_{2}}\left(z,t\right)$ by their mean fields $\left\langle\bar{\psi}_{P_{1}}\left(z,t\right)\right\rangle$ and $\left\langle\bar{\psi}_{P_{2}}\left(z,t\right)\right\rangle$ as well as neglected all terms quadratic in $\bar{\psi}_{S}\left(z,t\right)$ or higher.

\subsection{Solutions}\label{ssec:Solutions}
We note that equations for mean fields [Eqs.~\eqref{eq:psiSH},~\eqref{eq:psiP},~\eqref{eq:psiP1}, and~\eqref{eq:psiP2}]
can be found using standard approaches, such as split-step Fourier or finite difference methods~\cite{agrawal2007nonlinear}. The remaining operator equations [Eqs.~\eqref{eq:psiF},~\eqref{eq:dpsiP}, and~\eqref{eq:psiS}] are all of the form
\begin{align}
&\left(\frac{\partial}{\partial t}+v\frac{\partial}{\partial z}-i\frac{v^{\prime}}{2}\frac{\partial^{2}}{\partial z^{2}}\right)\bar{\psi}\left(z,t\right)\nonumber\\
&\qquad=\tilde{\mathcal{S}}\left(z,t\right)\bar{\psi}^{\dagger}\left(z,t\right)+2i\tilde{\mathcal{M}}\left(z,t\right)\bar{\psi}\left(z,t\right),
\label{eq:QFields}
\end{align}
where $\tilde{\mathcal{M}}\left(z,t\right)$ is real and positive.

Defining
\begin{align}
\omega\left(\kappa\right)&=v\kappa+\frac{v^\prime}{2}\kappa^{2},\nonumber\\
\tilde{\mathcal{S}}\left(z,t\right)&=\int\text{d}\kappa\frac{e^{i\kappa z}}{\sqrt{2\pi}}\mathcal{S}\left(\kappa,t\right),\nonumber\\
\tilde{\mathcal{M}}\left(z,t\right)&=\int\text{d}\kappa\frac{e^{i\kappa z}}{\sqrt{2\pi}}\mathcal{M}\left(\kappa,t\right),
\end{align}
where $\mathcal{M}\left(\kappa,t\right)=\mathcal{M}^{*}\left(-\kappa,t\right)$ and using [recall Eq.~\eqref{eq:psiz}]
\begin{equation}
\bar{\psi}\left(z,t\right)=\int\text{d}\kappa\frac{e^{i\kappa z}}{\sqrt{2\pi}}b\left(\kappa,t\right),
\end{equation}
where we have put $b_{J}\left(\kappa,t\right)=a_{J\left(k_{J}+\kappa\right)}\left(t\right)$, we can rewrite Eq.~\eqref{eq:QFields} as
\begin{align}
\left[\frac{\partial}{\partial t}+i\omega\left(\kappa\right)\right]b\left(\kappa,t\right)&=i\int\frac{\text{d}\kappa^{\prime}}{\sqrt{2\pi}}\mathcal{S}\left(\kappa+\kappa^{\prime},t\right)b^{\dagger}\left(\kappa^{\prime},t\right)\nonumber\\
&\quad+2i\int\frac{\text{d}\kappa^{\prime}}{\sqrt{2\pi}}\mathcal{M}\left(\kappa-\kappa^{\prime},t\right)b\left(\kappa^{\prime},t\right),
\label{eq:FTQFields}
\end{align}
a form highly amenable to numeric evaluation. Again we note that higher-order dispersion can be considered by including more terms in the expansion of the linear Hamiltonian of Eq.~\eqref{eq:HLpsi}, thus resulting in more terms in $\omega\left(\kappa\right)$.

To solve Eq.~\eqref{eq:FTQFields}, we discretize $\kappa_{j}=j\Delta\kappa$ and write
\begin{align}
&\int\frac{\text{d}\kappa^{\prime}}{\sqrt{2\pi}}\mathcal{S}\left(\kappa+\kappa^{\prime},t\right)b^{\dagger}\left(\kappa^{\prime},t\right)\nonumber\\
&\qquad\approx\sum_{j^{\prime}}\frac{\Delta\kappa}{\sqrt{2\pi}}\mathcal{S}\left(\kappa_{j}+\kappa_{j^{\prime}},t\right)b^{\dagger}\left(\kappa_{j^{\prime}},t\right),\nonumber\\
&\int\frac{\text{d}\kappa^{\prime}}{\sqrt{2\pi}}\mathcal{M}\left(\kappa-\kappa^{\prime},t\right)b\left(\kappa^{\prime},t\right)\nonumber\\
&\qquad\approx\sum_{j^{\prime}}\frac{\Delta\kappa}{\sqrt{2\pi}}\mathcal{M}\left(\kappa_{j}-\kappa_{j^{\prime}},t\right)b\left(\kappa_{j^{\prime}},t\right).
\end{align}
Introducing
\begin{equation}
\bm{Q}\left(t\right)=\left(\begin{array}{cc}
\bm{R}\left(t\right) & \bm{S}\left(t\right)\\
 -\bm{S}^{*}\left(t\right)  & -\bm{R}^{*}\left(t\right)
\end{array}\right),
\end{equation}
where
\begin{align}
R_{jj^{\prime}}\left(t\right)&=-\omega\left(\kappa_{j}\right)\delta_{jj^{\prime}}+2\frac{\Delta\kappa}{\sqrt{2\pi}}\mathcal{M}\left(\kappa_{j}-\kappa_{j^{\prime}},t\right),\nonumber\\
S_{jj^{\prime}}\left(t\right)&=\frac{\Delta\kappa}{\sqrt{2\pi}}\mathcal{S}\left(\kappa_{j}+\kappa_{j^{\prime}},t\right),
\end{align}
as well as the double vector
\begin{equation}
\bm{B}\left(t\right)=\left(\begin{array}{c}
\bm{b}\left(t\right)\\
\bm{b}^{\dagger}\left(t\right)
\end{array}\right), 
\end{equation}
where the vector $\bm{b}(t)$ has entries $b_j(t) = b(\kappa_j,t)$, 
then allows us to rewrite Eq.~\eqref{eq:FTQFields} compactly as
\begin{equation}
\frac{\partial}{\partial t}\bm{B}\left(t\right)=i\bm{Q}\left(t\right)\bm{B}\left(t\right).
\end{equation}
Note that the matrix $\bm{S}=\bm{S}^{T}$ is symmetric and $\bm{R}=\bm{R}^{\dagger}$ is Hermitian. For a small enough propagation forward in time $\Delta t$ this has solution
\begin{equation}\label{eq:FwdFields}
\bm{B}\left(t+\Delta t\right)=\bm{K}(t)\bm{B}\left(t\right),
\end{equation}
where the (single-time) infinitesimal propagator is defined as
\begin{equation}
\bm{K}(t)=\exp\left[i\Delta t\bm{Q}\left(t\right)\right]=\left(\begin{array}{cc}
\bm{V}(t) & \bm{W}(t)\\
 \bm{W}^{*}(t) & \bm{V}^{*}(t) 
\end{array}\right).
\end{equation}
Propagation over a finite interval can then be calculated by concatenating infinitesimal propagators $\bm{K}(t)$ to form the (two-argument) Heisenberg picture propagator~\cite{quesada2019efficient} 
\begin{align}\label{eq:cron}
\bm{K}(t,t_0) = \prod_{j=1}^\ell \bm{K}(t_j) = \left(\begin{array}{cc}
\bm{V}(t,t_0) & \bm{W}(t,t_0)\\
\bm{W}^{*}(t,t_0) & \bm{V}^{*}(t,t_0)
\end{array}\right),
\end{align}
where $t_j = t_0+j \Delta t$ and $\Delta t = (t-t_0)/\ell$,
as
\begin{align}
\bm{B}\left(t \right) = \bm{K}(t,t_0) \bm{B}\left(t_0 \right),
\end{align}
and the product in Eq.~\eqref{eq:cron} is taken in chronological order.
Reverting back to continuous notation, we note that we may also write the solution of Eq.~\eqref{eq:FTQFields} as 
\begin{align}\label{eq:heissol}
b(\kappa,t) =& \int \dd \kappa'\, \mathcal{V}(\kappa,\kappa';t,t_0) b(\kappa',t_0)\\
&+ \int \dd \kappa'\, \mathcal{W}(\kappa,\kappa';t,t_0) b^\dagger(\kappa',t_0), \nonumber
\end{align}
where 
\begin{subequations}\label{eq:conttodisc}
\begin{align}
\mathcal{V}(\kappa_j,\kappa_{j'};t,t_0) &= \bm{V}_{j,j'}(t,t_0)/\Delta \kappa,\\
\mathcal{W}(\kappa_j,\kappa_{j'};t,t_0) &= \bm{W}_{j,j'}(t,t_0)/\Delta \kappa.
\end{align}
\end{subequations}
We note that similar input-output relations have been derived previously \cite{mckinstrie2013quadrature, christ2013theory,horoshko2017bloch,kolobov1999spatial}, albeit with different labels. 
\subsection{Including loss}
\label{ssec:loss}
For unitary evolution (i.e.\ ignoring scattering loss), the evolution of the $b\left(\kappa,t\right)$ themselves is sufficient to characterize system dynamics. However, once we include loss it becomes extremely useful to consider the phase insensitive and phase sensitive second moments with entries
\begin{align}\label{eq:MomentsDef}
N_{i,j}\left(t\right)&=\left\langle b_i^{\dagger}\left(t\right)b_j\left(t\right)\right\rangle, \nonumber \\
M_{i,j}\left(t\right)&=\left\langle b_i\left(t\right)b_j\left(t\right)\right\rangle.
\end{align}
Using Eq.~\eqref{eq:FwdFields} one can easily write equations that update these moments as
\begin{align}\label{eq:FieldsUpdateUnitary}
\bm{N}(t+\Delta t) &= \bm{W}^* \bm{M}(t) \bm{V}^T+\bm{V}^* \bm{M}^*(t) \bm{W}^T\nonumber\\
&\quad+\bm{V}^* \bm{N}(t) \bm{V}^T+\bm{W}^* \bm{N}^T(t) \bm{W}^T\nonumber\\ &\quad+\bm{W}^*\bm{W}^T,\nonumber\\
\bm{M}(t+\Delta t) &= \bm{V} \bm{M}(t) \bm{V}^T+\bm{W}\bm{M^*}(t) \bm{W}^T\nonumber\\
&\quad+\bm{W}\bm{N}(t)\bm{V}^T + \bm{V}\bm{N}^T(t) \bm{W}^T\nonumber\\
&\quad+\bm{V} \bm{W}^T.
\end{align}
where $\bm{V} \equiv \bm{V}(t)$ and $\bm{W} \equiv \bm{W}(t)$ are the (single-argument) blocks of the propagator $\bm{K}(t)$.
Note that in these equations there are inhomogeneous terms that drive vacuum fluctuations into the system and cause the moments to acquire nonzero values.

To include loss in a Hamiltonian formalism we need to add extra degrees of freedom to the system that will account for the modes into which the lost photons go. To this end we consider $t$ as the time the waveguide modes interact with non-guided modes $c(\kappa,t)$ that are initially in vacuum via a hopping interaction that removes photons from the $b(\kappa,t)$ and places them in the $c(\kappa,t)$. The operators of the system will transform as
\begin{align}
b(\kappa,t+\Delta t) = \sqrt{\eta} b(\kappa,t) + \sqrt{1-\eta} \ c(\kappa,t).
\end{align}
For the moments one finds
\begin{align}\label{eq:FieldsUpdateLoss}
\bm{N}(t+\Delta t) = \eta \bm{N}(t), \quad \bm{M}(t+\Delta t) = \eta \bm{M}(t),
\end{align}
where recall that at time $t$ the scattering modes $c$ are in vacuum. This recipe allows us to update the moments of the waveguide after experiencing a small amount of loss in the time interval $\Delta t$. For larger time intervals, we simply concatenate Eq.~\eqref{eq:FieldsUpdateLoss} for many $\Delta t$s. The key assumption for the validity of this evolution is that at each time interval $\Delta t$ the system of interest interacts with ``fresh'' non-guided modes that always come prepared in vacuum. This is essentially the same Markov approximation used to derive the Lindblad form Master Equation~\cite{fischer2018derivation}. 

The only remaining question is how to set the value of $\eta$. Since we expect loss to compound exponentially it is natural to take 
\begin{align}\label{eq:expdecay}
\eta = \exp(-\gamma \Delta t),
\end{align}
where $\gamma$ can be easily obtained from the standard attenuation constant $\alpha$~\cite{agrawal2007nonlinear}. 
Numerically, combining the rules in Eq.~\eqref{eq:FieldsUpdateUnitary} and Eq.~\eqref{eq:FieldsUpdateLoss} one can propagate the moments of the state (cf. Appendix \ref{sec:momentproc} for a detailed derivation). Note that, because both squeezing and loss are Gaussian operations, if the state in the waveguide is Gaussian at the beginning of the propagation it will remain Gaussian throughout the evolution and thus be fully characterized by the $\bm{N}$ and $\bm{M}$ moments as well as the expectation value $\braket{b(\kappa,t)}$. 

Finally, it is easy to show, after some algebra, that if the state at $t_0$ is $\ket{\Psi(t_0)} = \ket{\text{vac}}$ then the state at some later time $t$ is the squeezed state
\begin{align}
&\ket{\Psi(t)} = \\
&\exp\left(\int \dd{k} \dd{k'}\, J(k,k';t,t_0) b^\dagger(k,t_0) b^\dagger(k,t_0) -\text{H.c.} \right) \ket{\text{vac}}, \nonumber
\end{align}
where the joint spectral amplitude
\begin{align}\label{eq:jsadef}
 J(k,k';t,t_0) = \sum_{l} r_{l}\ \rho^{(l)}(k)\rho^{(l)}(k'),
\end{align}
is written in terms of the Schmidt functions and coefficients of the phase sensitive moment
\begin{align}
&\braket{\text{vac}|b^{}(k,t)b^{}(k',t)|\text{vac}} \nonumber\\
& \quad =\int \dd{k''}\,\mathcal{V}(k,k'';t,t_0)\mathcal{W}^{*}(k',k'',t,t_0)\nonumber \\
& \quad =\sum_{l}\frac{\sinh(2 r_{l})}{2}\rho^{(l)}(k)\rho^{(l)}(k').
\end{align}
This is because the Hamiltonian generating $\mathcal{U}(t,t_0)$ [cf.\ Eq.~\eqref{eq:defU}], even though  nonlinear in the fields, is, upon replacing the pump(s) by its (their) classical expectation value(s), quadratic in the quantum bosonic operators~\cite{serafini2017quantum, weedbrook2012gaussian}.

The expressions developed here constitute our main results. Starting from an equation of the form of Eq.~\eqref{eq:QFields}, describing pulsed pump SPDC or SFWM, they show how to include dispersion to any order, as well as SPM, XPM, and loss, in determination of the output state, its joint spectral amplitude, and phase sensitive and phase insensitive moments for arbitrary input pump power. These results provide complete information about the degenerate squeezed state of light generated in a waveguide. In what follows, we provide examples of their utility.

\section{Example calculations}
\label{sec:examples}
\subsection{SPDC in the low-gain regime}
As a first application of our formalism, we consider SPDC in the low-gain regime, where results are well known and can be obtained analytically~\cite{mosley2008heralded}. This serves as a basic check on the results of the previous section, and provides a simple rule of thumb for the grid size one should use to represent the moments when solving the problem numerically.

In Appendix \ref{sec:momentproc} we derive the continuous evolution of the moments in this regime including loss. Solving the equations of motion between $t_0 = -T/2$ and $t_1= T/2$ perturbatively  we find
\begin{align}
&\braket{b_{F}(\kappa,t_1)b_{F}(\kappa',t_1)} = i \int \frac{\dd \kappa''}{(2 \pi)^{3/2}} e^{i \mu}   \braket{b_{SH}(\kappa'',t_0)} \label{eq:SPDCmoment} \\
& \quad \quad \times \Phi(\kappa+\kappa'-\kappa'')  \exp\left( i\epsilon^{+} T/2 \right) \frac{\sinh\left(i T \epsilon^{-} /2 \right)}{i\epsilon^{-}/2}, \nonumber
\end{align}
where we have put
\begin{align}
\mathcal{\tilde{S}}(z,t) &= \zeta^{\left(2\right)}(z) \braket{\bar{\psi}_{SH}(z,t)},\\
\braket{\bar{\psi}_{SH}(z,t)} &= \int \frac{\dd \kappa}{\sqrt{2 \pi}} \braket{b_{SH}(\kappa,t_0)}  \label{eq:pumpevol}\\
& \quad\quad  \quad  \quad \times  e^{i \kappa z} e^{-i(\omega_{SH}(\kappa) -i \gamma_{SH}/2) (t-t_0)} , \nonumber \\
\mu &= T\left\{\omega_F(\kappa)+\omega_F(\kappa')+\omega_{SH}(\kappa'')\right\},
\end{align}
and assumed that the pump mean field undergoes only free propagation.
Furthermore, we have introduced the phase-matching function
\begin{align}\label{eq:pmf}
\Phi(\kappa+\kappa'-\kappa'') \equiv \int \dd z \  \zeta^{\left(2\right)}(z) e^{-i(\kappa+\kappa'-\kappa'')z},
\end{align}
and the complex energy difference and sum
\begin{align}\label{eq:cenergies}
\epsilon^{\pm} =  \omega_F(\kappa)+\omega_F(\kappa') \pm \omega_{SH}(\kappa'') \pm i \frac{\gamma_F+\gamma_F\pm\gamma_{SH}}{2}.
\end{align}
Note that the loss coefficients appear in two different places in the expression for $\braket{b_{F}(\kappa,t_1)b_{F}(\kappa',t_1)}$: they appear as a sum $\epsilon^{+}$ in an overall exponential damping factor, but, more importantly, they also appear as a difference $\epsilon^{-}$ in the $\sinh$ loss matching term~\cite{antonosyan2014effect, helt2015spontaneous}. 

When loss is negligible one can let $T \to \infty$ and write 
\begin{align}
\frac{\sinh\left( i T \epsilon^{-}/2 \right)}{\epsilon^{-}/2} =  \frac{\sin\left( \Delta \omega \tfrac{T}{2}\right)}{\Delta  \omega/2 } = 2\pi \delta( \Delta \omega),% \text{ when } T\to\infty
\end{align}
where $\Delta \omega= \omega_F(\kappa) +\omega_F(\kappa') - \omega_P(\kappa'')$. With this simplification one can rewrite Eq.~\eqref{eq:SPDCmoment} as
\begin{align}
&\braket{b_{F}(\kappa,t_1)b_{F}(\kappa^{\prime},t_1)}  \nonumber \\
&=i \int \frac{\dd \kappa''}{\sqrt{2 \pi}} e^{i \mu}   \braket{b_{SH}(\kappa'',t_0)} \Phi(\kappa+\kappa'-\kappa'')  \delta(\Delta \omega),
\end{align}
dropping an irrelevant phase factor. Finally, moving to frequency space yields the well known expression
\begin{align}
&\braket{b_F(\omega,t_1) b_F(\omega',t_1)}= (2 \pi v_F v_F v_{SH})^{-1/2} \\
&\times \braket{b_{SH}(\omega+\omega',t_0)} \Phi(\kappa_F(\omega) + \kappa_F(\omega')-\kappa_{SH}(\omega+\omega')), \nonumber
\end{align}
showing that the low gain JSA (recall that the JSA and the phase sensitive moment are equal in the low gain regime) is a product of the pump and phase-matching functions.

\begin{figure}
	\begin{center}
		\includegraphics[width=0.95\columnwidth]{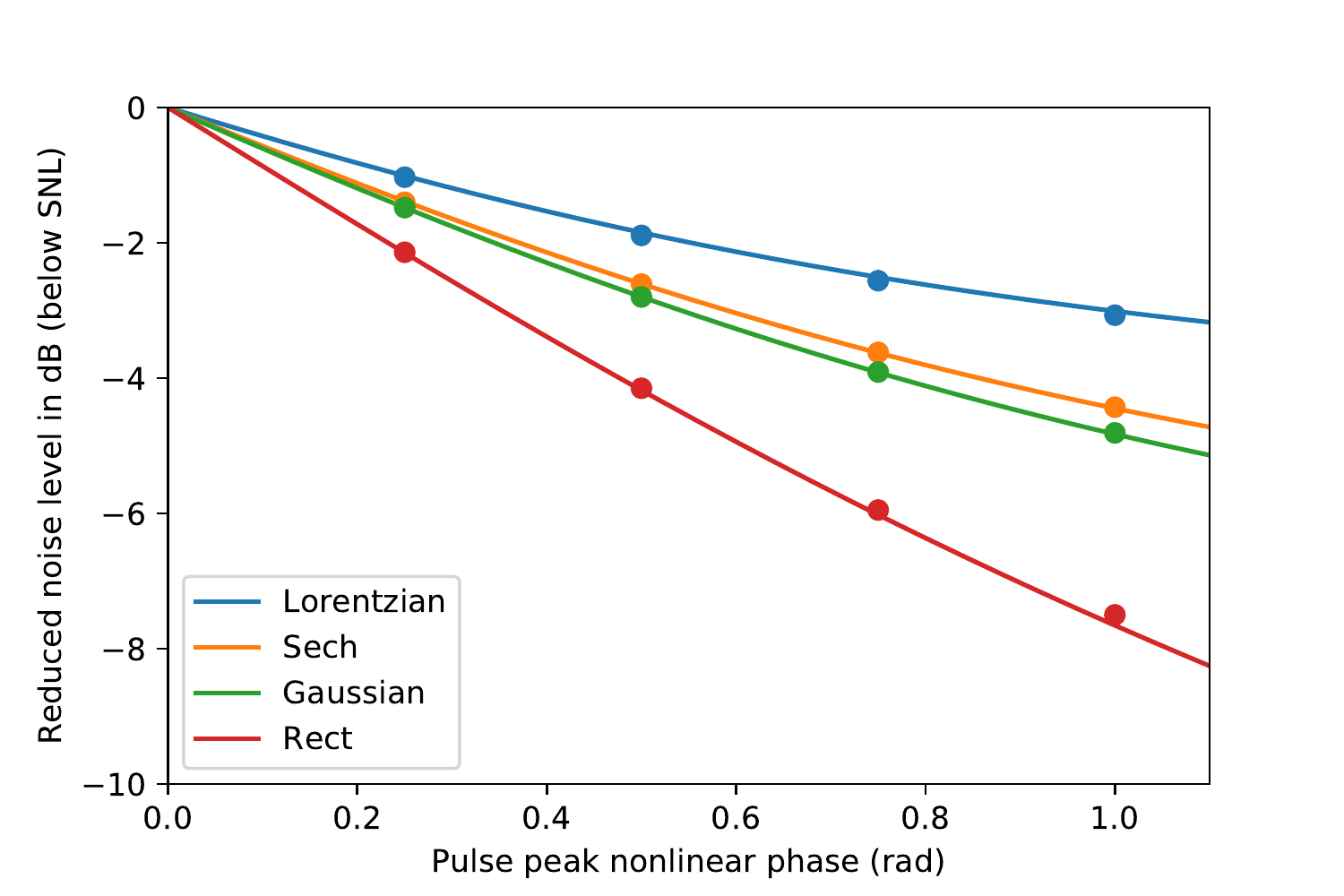}
	\end{center}
	\caption{Squeezing below the shot noise level (SNL). The solid lines represent analytic results obtained with Eq.~\eqref{eq:analytichomodyne} and the individual points results obtained with Eqs.~\eqref{eq:MomentsDef} and \eqref{eq:discreteV}. Neither case includes loss or second-order dispersion.}
	\label{fig:homodyne}
\end{figure}
\subsection{Single-pump SFWM quadrature variance}
As a second check, we move away from analytic results and the perturbative regime, but neglect both loss and dispersion. We consider a homodyne measurement, in which one measures $V_{\theta}=\left\langle X_{\theta}^{2}\right\rangle -\left\langle X_{\theta}\vphantom{X_{\theta}^{2}}\right\rangle ^{2}$ where
\begin{equation}
X_{\theta}=e^{i\theta}\mathcal{B}+e^{-i\theta}\mathcal{B}^{\dagger},
\end{equation}
with
\begin{equation}
\mathcal{B}=\int\text{d}\kappa\,\phi^{*}\left(\kappa\right)b\left(\kappa,t_{f}\right),
\end{equation}
for a normalized function $\int\text{d}\kappa\,\left|\phi\left(\kappa\right)\right|^{2}=1$,
and ${t_{f}=L/v}$ the time it takes light to traverse the length of the nonlinear region $L$. Writing this out in detail, we find
\begin{align}
V_{\theta}&=e^{2i\theta}\int\text{d}\kappa\text{d}\kappa^{\prime}\,\phi^{*}\left(\kappa\right)\phi^{*}\left(\kappa^{\prime}\right)\left\langle b\left(\kappa,t_{f}\right)b\left(\kappa^{\prime},t_{f}\right)\right\rangle +\text{c.c.}\nonumber\\
&\quad+2\int\text{d}\kappa\text{d}\kappa^{\prime}\,\phi\left(\kappa\right)\phi^{*}\left(\kappa^{\prime}\right)\left\langle b^{\dagger}\left(\kappa,t_{f}\right)b\left(\kappa^{\prime},t_{f}\right)\right\rangle+1,
\label{eq:Vtheta}
\end{align}
where the final term sets the so-called shot noise level. We note that when $\phi\left(\kappa\right)$ corresponds to a Schmidt function $\rho^{\left(l\right)}\left(\kappa\right)$ [recall Eq.~\eqref{eq:jsadef}], maximizing and minimizing $\left(\pm\right)$ over $\theta$, this reduces to the more common
\begin{align}
V^{\pm}&=2\sinh^{2}\left(r_{l}\right)\pm\sinh\left(2r_{l}\right)+1\nonumber\\
&=e^{\pm2r_{l}}.
\end{align}
For numeric evaluation, we discretize $\phi_{j}=\phi\left(\kappa_{j}\right)\sqrt{\Delta\kappa}$ and rewrite Eq.~\eqref{eq:Vtheta} as
\begin{equation}
V_{\theta}=e^{2i\theta}\bm{\phi}^{*}\bm{M}\left(t_{f}\right)\bm{\phi}^{\dagger}+\text{c.c.}+2\bm{\phi}\bm{N}\left(t_{f}\right)\bm{\phi}^{\dagger}+1.
\label{eq:discreteV}
\end{equation}

We can check results obtained using Eq.~\eqref{eq:discreteV} in the limit of no loss and no second-order dispersion (i.e. ${v'=0}$), as the mean field equation for single-pump SFWM [Eq.~\eqref{eq:psiP}] can be solved analytically in a moving frame travelling with velocity $v_{P}$ such that $t\rightarrow t-z/v_{P}t$,
\begin{equation}
\frac{\partial}{\partial z}\left\langle \bar{\psi}_{P}\left(z,t\right)\right\rangle =i\frac{\zeta_{PPPP}^{\left(3\right)}\left(z\right)}{v_{P}}\left|\left\langle \bar{\psi}_{P}\left(z,t\right)\right\rangle \right|^{2}\left\langle \bar{\psi}_{P}\left(z,t\right)\right\rangle.
\end{equation}
Ignoring the $z$ dependence of $\zeta_{PPPP}^{\left(3\right)}$ and putting $\sqrt{\hbar\omega_{P}v_{P}}\left\langle \bar{\psi}_{P}\left(z,t\right)\right\rangle\rightarrow A_{P}\left(z,t\right)$ such that $\left\vert A_{P}\left(z,t\right)\right\vert^{2}$ has units of power, we find
\begin{equation}
\frac{\partial}{\partial z}A\left(z,t\right)=i\gamma\left|A\left(z,t\right)\right|^{2}A\left(z,t\right),
\end{equation}
with solution
\begin{equation}
A\left(z,t\right)=e^{i\gamma\left|A\left(0,t\right)\right|^{2}z}A\left(0,t\right),
\end{equation}
where we have introduced the nonlinear parameter $\gamma$ (see Appendix \ref{app:CMEs})~\cite{agrawal2007nonlinear}. Following Ref.~\cite{shirasaki1990squeezing} (cf. Eq.~(5.8a) there) the quadrature variance minimum can then be written as
\begin{align}
\label{eq:analytichomodyne}
&V^{-}=\frac{1}{\int\text{d}t\,\Phi\left(t\right)}\\
&\times\int\text{d}t\,\Phi\left(t\right)\left[1+2\Phi^{2}\left(t\right)-\frac{2\Phi\left(t\right)\left[1+\Phi\left(t\right)C\Phi\left(0\right)\right]}{\sqrt{1+C^{2}\Phi^{2}\left(0\right)}}\right], \nonumber
\end{align}
where $\Phi\left(t\right)=\left|A\left(0,t\right)\right|^{2}\gamma L$ is assumed peaked at $t=0$ and $C$ is a constant specific to pump pulse shape. In Fig.~\ref{fig:homodyne} we plot this analytic $V^{-}$ (solid lines) as well as results from our numerical routines (points) in dB as functions of $\Phi\left(0\right)$ for Lorentzian ($C=3/4$), Sech ($C=4/5$), Gaussian ($C=\sqrt{2/3}$), and Rectangular ($C=1$) pulse shapes.

\begin{figure}
	\includegraphics[width=1.05\columnwidth]{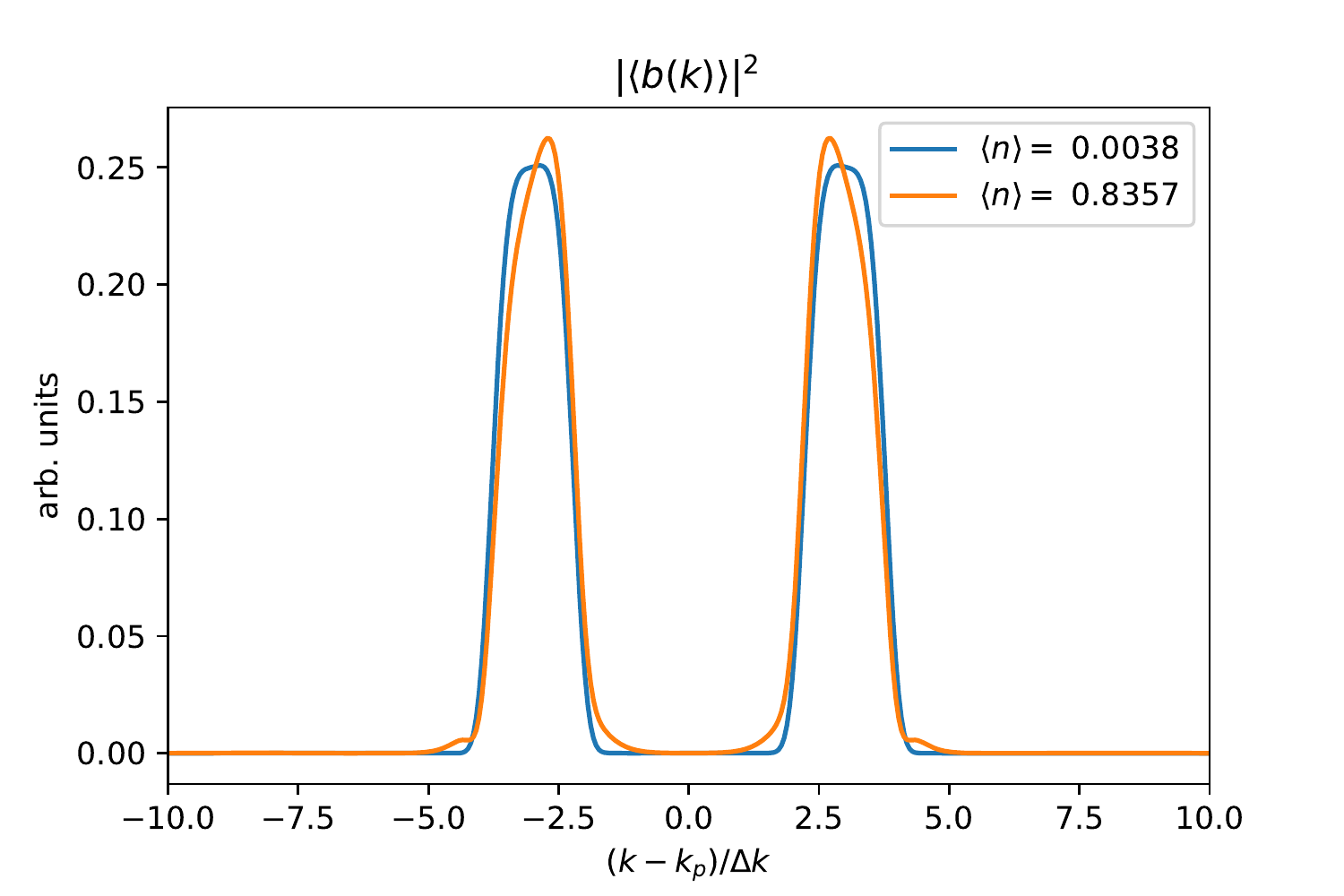}
	\caption{\label{fig:mean_density}$L^2$ normalized photon density of the mean field after propagating through the nonlinear region. Note how its shape remains relatively similar to the input for a mean photon number of the squeezed fluctuations $\langle n \rangle \ll 1$, but changes shape due to the effects of SPM for large enough $\langle n \rangle$.}
\end{figure}
\begin{figure}
	\includegraphics[width=1.05\columnwidth]{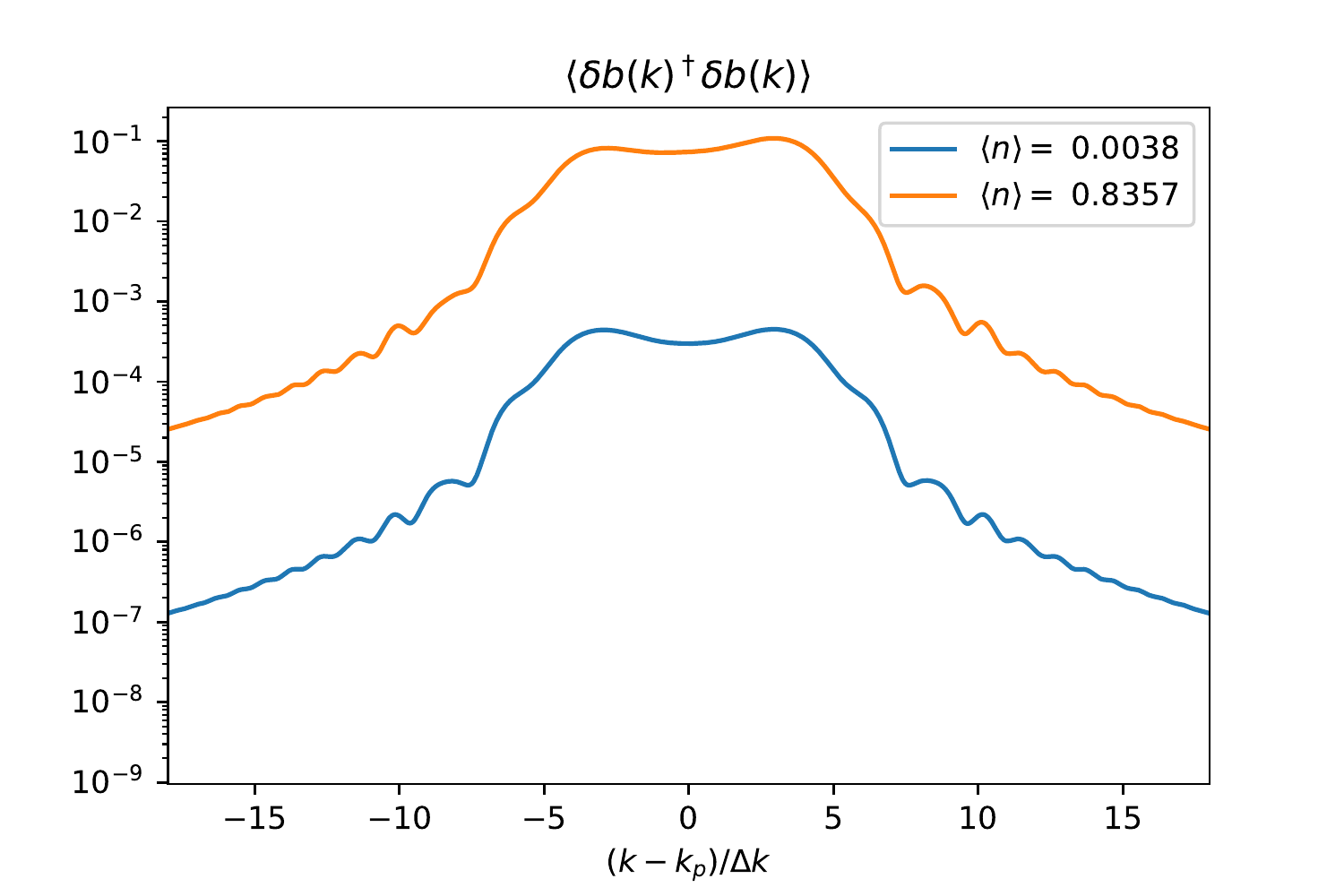}
	\caption{\label{fig:squeezed_density} Photon density of the fluctuations after propagating through the nonlinear region. The shape is quite similar for both a mean photon number of the fluctuations $\langle n \rangle = 0.0038$ and $\langle n \rangle = 0.8357$, though note that it is becoming slightly asymmetric, due to the interplay of dispersion and XPM, for $\langle n \rangle = 0.8357$. }
\end{figure}

\begin{figure*}
	\includegraphics[width=0.7\textwidth]{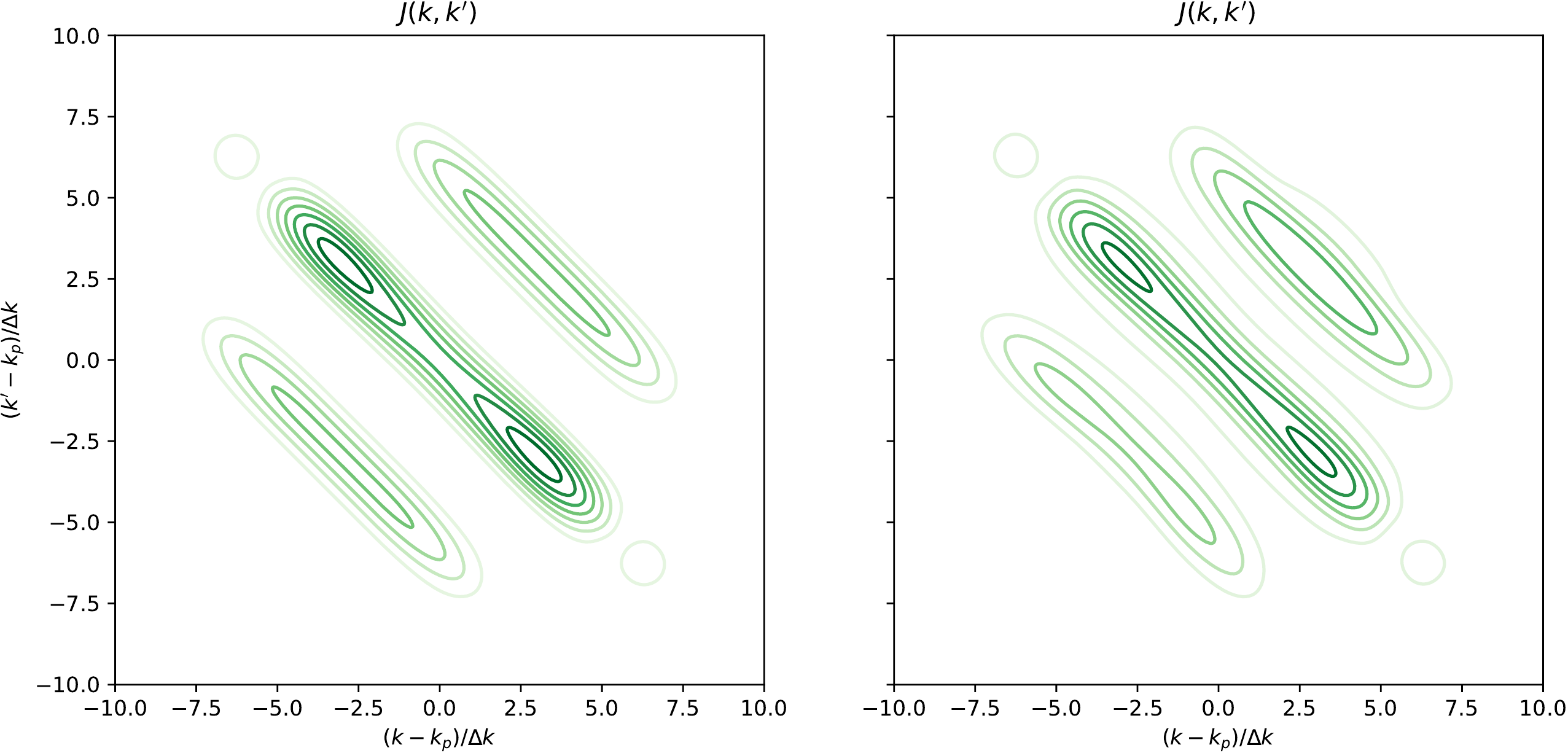}
	\caption{\label{fig:jsa}Joint spectral amplitude $J(k,k')$ for different gains specified by the mean photon number $\langle n \rangle$ of the squeezed fluctuations of the field. Note the change in the shape of the JSA as  $\langle n \rangle$ increases from 0.0038 to 0.8357.}
\end{figure*}

\begin{figure}
	\includegraphics[width=0.40\textwidth]{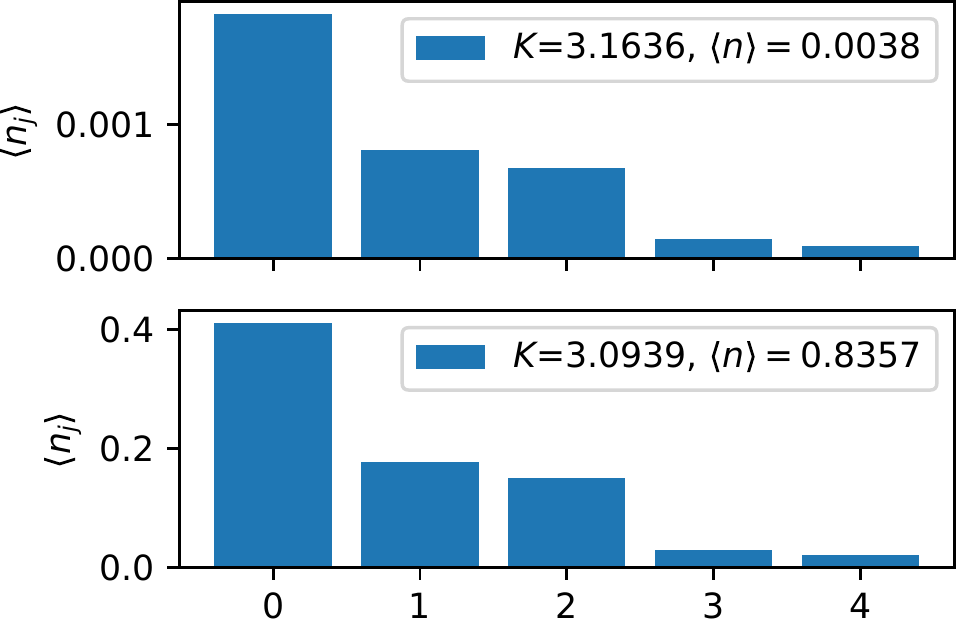}
	\caption{\label{fig:schmidt}Distribution of Schmidt modes for mean photon number of the fluctuations $\langle n \rangle = 0.0038$ and $\langle n \rangle = 0.8357$. Note how the Schmidt number is reduced as $\langle n \rangle$ is increased and a larger fraction of the total photons are generated in the lowest order Schmidt mode. }
\end{figure}
\subsection{Dual-pump SFWM in the high-gain regime}
Finally, we consider dual-pump SFWM in the high-gain regime, the situation that is most relevant for GBS, as achieving degenerate squeezing via SPDC can require significant dispersion engineering, and single-pump SFWM demands interferometric removal of the pump from the generated squeezed state. As noted earlier~\cite{vernon2018scalable}, an ideal degenerate squeezing source for GBS applications should satisfy four desiderata, namely:
(i) scalability, (ii) single-mode operation, (iii) squeezing levels sufficient to enable a genuine quantum advantage in computation, and (iv) compatibility with single photon and photon number-resolving detection.
We investigate the suitability of a dual pump source with respect to (ii-iv). We note that integrated platforms are some of the best suited platforms for scalability, given their ease of fabrication and integration with other optical components.

We consider a silicon waveguide where one aims to generate squeezing around 1550 nm (corresponding to $k \approx 2\pi / 1550 \text{ nm}$). For simplicity, we neglect linear and nonlinear loss, noting that adding such decoherence mechanisms will only degrade the quality of the source. We take the two pumps to each have a bandwidth $\Delta k = 41.8$ cm$^{-1}$ (corresponding to $\Delta \lambda \approx 1.6$ nm), and have centres separated by approximately $3 \Delta k$, with shapes similar to a top hat function (this corresponds roughly to the parameters used in Ref.~\cite{paesani2019generation}). Note that it is inappropriate to describe the pumps and signal field as three distinct modes in this situation, as the wavevector support of the pumps and the generated squeezed light significantly overlap and thus it does not makes sense to attach different mode labels to the pump and squeezed light. Based on this observation we directly use the theory developed to propagate mean values [cf.~Eq.~\eqref{eq:psiP}] and fluctuations [cf.~Eq.~\eqref{eq:dpsiP}] for single-pump SFWM, elaborated upon in Sec.~\ref{ssec:Solutions}. The pump has a bimodal distribution in its wavevectors (or frequencies) justifying the use of the term dual-pump.

Before entering the nonlinear region the field is prepared in a coherent state
with amplitude
\begin{align}
\braket{b(k,t_0)} =&   \sqrt{\tfrac{N}{C}} \times\\
& \left\{ \exp\left(-\left(\tfrac{k-k_1}{\Delta k} \right)^4\right)+\exp\left(-\left(\tfrac{k-k_2}{\Delta k} \right)^4\right) \right\}, \nonumber 
\end{align}
where we have used the function $f(k) = e^{-k^4}$ as a smooth non-Gaussian approximation to a top-hat function of unit width, $k_2 - k_1 = 3 \Delta k$, $\tfrac{k_1+k_2}{2} = k_P = \frac{2 \pi}{1550 \text{ nm}}$, $N$ is the mean photon number in the pumps, and $C$ is a normalization constant.
We truncate the dispersion relation at the quadratic level as in  Eq.~\eqref{eq:disprel} with the parameters $v_P = 7.019 \times 10^{7}$ m/s and $v_P'=4.711$ m$^2$/s. 

In Fig.~\ref{fig:mean_density} we show the photon density of the mean field for different values of the gain. We parametrize this gain in terms of the mean photon number of the squeezed fluctuations generated in the field
\begin{align}
\braket{n} &= \sum_{j} \braket{n_j} = \int dk \  \mathcal{N}(k), \\
\mathcal{N}(k)&= \braket{\delta b^\dagger(k) \  \delta b(k)} \label{eq:sqphden},
\end{align}
where $\braket{n_j} = \sinh^2 r_j$ is the mean photon number in the $j^{\text{th}}$ Schmidt mode and we have written the destruction operator at wavevector $k$ as its mean plus its fluctuations $b(k) = \braket{b(k)}+\delta b(k)$. The photon density in the mean can be contrasted with the photon density in the squeezed fluctuations defined in Eq.~\eqref{eq:sqphden} and shown in Fig.~\ref{fig:squeezed_density}. Note that there is significant overlap between the mean and the fluctuations; filtering is thus necessary to prevent the large amounts of energy in the mean from blinding the detectors necessary for applications in quantum sampling. Note, however, that the filtering of non-classical light comes with its own complications. In particular this will introduce mixedness in the state; this has been investigated in twin-beam sources in~\cite{blay2017effects,branczyk2010optimized,meyer2017limits} and will be explored for degenerate squeezing in a future communication.
The problem of overlap between the field and the squeezed fluctuations is further compounded by the fact that the joint spectral amplitude [cf.~Eq.~\eqref{eq:jsadef}] of the squeezed fluctuations is spectrally non-separable. This can be seen by looking at its absolute value for different gains as shown in Fig.~\ref{fig:jsa}. 
We numerically confirm that the JSAs are not separable by calculating the Schmidt number, defined as
\begin{align}
K = \frac{\left( \sum_j \braket{n_j} \right)^2}{\sum_j \braket{n_j}^2}.
\end{align}
This quantity takes the value 1 for a separable JSA and increases for more entangled JSAs. For the source we model we find $K \sim 3.1 $ both in the low- and high-gain regimes (cf.\ Fig.~\ref{fig:schmidt}).  Comparing against recently proposed degenerate squeezing sources using microrings, for which $K \sim 1.0$ \cite{vernon2018scalable}, suggests that sources with parameters similar to those we have considered in this section are ill-suited for quantum sampling applications, and that the parameter space should be explored in a more thorough way to find better operating configurations.

\section{Conclusions}
\label{sec:conclusions}
In conclusion, we have presented a unified Hamiltonian formalism for squeezed states of light generated via either SPDC or SFWM in waveguides using pulsed pumps. Our straightforward techniques directly admit dispersion to any order, linear losses, and self- and cross-phase modulation. They can be used to calculate standard expectation values, such as moments and spectra, but also states (including joint spectral amplitudes) and thus any desired observables. 

We have provided examples of the application of our formalism to the well-known examples of SPDC in the low-gain regime and homodyne  detection of single-pump SFWM, demonstrating its utility. All of the Python code used in this work is available on GitHub~\cite{code}. Furthermore, we have explored the suitability of a standard waveguide dual-pump SFWM source of squeezed light for GBS applications, finding that it is difficult to obtain a Schmidt number near unity in such a source. This points to a need to continue to explore more exotic and engineered sources going forward, a task for which our formalism is suited.

\begin{acknowledgments}
The authors thank M. Menotti for waveguide mode simulations as well as M. J. Steel, G. Triginer, V. D. Vaidya, and J. E. Sipe for helpful discussions.
\end{acknowledgments}

\appendix
\section{Derivation of coupled-mode equations}
\label{app:CMEs}
\subsection{SPDC}
Our second-order nonlinear coupling parameter is given by
\begin{align}
\zeta^{\left(2\right)}\left(z\right)	&=\frac{2}{\hbar}\frac{1}{\varepsilon_{0}}\sqrt{\frac{\left(\hbar\bar{\omega}_{F}\right)^{2}\left(\hbar\bar{\omega}_{SH}\right)}{2^{3}}}\int\dd{x}\dd{y}\,\left[\Gamma_{jlm}^{\left(2\right)}\left(\bm{r}\right)\vphantom{\left[d_{Fk_{F}}^{j}\left(x,y\right)d_{Fk_{F}}^{l}\left(x,y\right)\right]^{*}}\right.\nonumber\\
&\quad\times\left.\left[d_{Fk_{F}}^{j}\left(x,y\right)d_{Fk_{F}}^{l}\left(x,y\right)\right]^{*}d_{SHk_{SH}}^{m}\left(x,y\right)\right],
\end{align}
or, putting
\begin{equation}
\Gamma_{jlm}^{\left(2\right)}\left(\bm{r}\right)\rightarrow\frac{\chi_{jlm}^{\left(2\right)}\left(x,y\right)s\left(z\right)}{\varepsilon_{0}n^{4}\left(x,y;\bar{\omega}_{F}\right)n^{2}\left(x,y;\bar{\omega}_{SH}\right)},
\end{equation}
\begin{equation}
\zeta^{\left(2\right)}\left(z\right)	=\frac{2}{\hbar}\sqrt{\frac{\left(\hbar\bar{\omega}_{F}\right)^{2}\left(\hbar\bar{\omega}_{SH}\right)}{2^{3}\varepsilon_{0}\mathcal{A}^{\left(2\right)}}}\frac{s\left(z\right)\bar{\chi}^{\left(2\right)}}{\bar{n}^{3}},
\end{equation}
where we have introduced
\begin{align}
\mathcal{A}^{\left(2\right)}&	=\left|\int\text{d}x\text{d}y\frac{\chi_{jlm}^{\left(2\right)}\left(x,y\right)\bar{n}^{3}}{\varepsilon_{0}^{3/2}\bar{\chi}^{\left(2\right)}}\vphantom{\frac{\left[d_{Fk_{F}}^{j}\left(x,y\right)\right]^{*}}{n^{4}\left(x,y;\omega_{F}\right)}}\right.\nonumber\\
&\quad\times\left.\frac{\left[d_{Fk_{F}}^{j}\left(x,y\right)d_{Fk_{F}}^{l}\left(x,y\right)\right]^{*}d_{SHk_{SH}}^{m}\left(x,y\right)}{n^{4}\left(x,y;\bar{\omega}_{F}\right)n^{2}\left(x,y;\bar{\omega}_{SH}\right)}\right|^{-2}.
\end{align}
Here $s\left(z\right)$ defines the nonlinear region of the waveguide, and $\bar{n}$ and $\bar{\chi}^{2}$ are, respectively, a refractive index and second-order susceptibility introduced solely for convenience. The general operator coupled-mode equations resulting from Eqs.~\eqref{eq:HLpsi} and~\eqref{eq:HSPDC} of the main text are 
\begin{align}
&\left(\frac{\partial}{\partial t}+v_{SH}\frac{\partial}{\partial z}-i\frac{v_{SH}^{\prime}}{2}\frac{\partial^{2}}{\partial z^{2}}\right)\bar{\psi}_{SH}\left(z,t\right)\nonumber\\
&\qquad=\frac{i}{2}\left[\zeta^{\left(2\right)}\left(z\right)\right]^{*}\bar{\psi}_{F}\left(z,t\right)\bar{\psi}_{F}\left(z,t\right),
\end{align}
and
\begin{align}
&\left(\frac{\partial}{\partial t}+v_{F}\frac{\partial}{\partial z}-i\frac{v_{F}^{\prime}}{2}\frac{\partial^{2}}{\partial z^{2}}\right)\bar{\psi}_{F}\left(z,t\right)\nonumber\\
&\qquad=i\zeta^{\left(2\right)}\left(z\right)\bar{\psi}_{F}^{\dagger}\left(z,t\right)\bar{\psi}_{SH}\left(z,t\right).
\end{align}
However, replacing $\bar{\psi}_{SH}\left(z,t\right)$ by its mean field $\bar{\psi}_{SH}\left(z,t\right)\rightarrow\left\langle\bar{\psi}_{SH}\left(z,t\right)\right\rangle$ and neglecting terms quadratic in $\bar{\psi}_{F}\left(z,t\right)$ leads to Eqs.~\eqref{eq:psiSH} and~\eqref{eq:psiF} of the main text.

\subsection{Single-pump SFWM}
Our third-order nonlinear coupling parameter is given by
\begin{align}
\zeta_{J_{1}J_{2}J_{3}J_{4}}^{\left(3\right)}\left(z\right)&=\frac{2}{\hbar}\frac{3}{2\varepsilon_{0}}\sqrt{\frac{\left(\hbar\bar{\omega}_{J_{1}}\right)\left(\hbar\bar{\omega}_{J_{2}}\right)\left(\hbar\bar{\omega}_{J_{3}}\right)\left(\hbar\bar{\omega}_{J_{4}}\right)}{2^{4}}}\nonumber\\
&\quad\times\int\dd{x}\dd{y}\,\left[\Gamma_{jlmn}^{\left(3\right)}\left(\bm{r}\right)\vphantom{\left[d_{J_{1}k_{J_{1}}}^{j}\left(x,y\right)d_{J_{2}k_{J_{2}}}^{l}\left(x,y\right)\right]^{*}}\left[d_{J_{1}k_{J_{1}}}^{j}\left(x,y\right)\right]^{*}\right.\nonumber\\
&\quad\times\left.\left[d_{J_{2}k_{J_{2}}}^{l}\left(x,y\right)\right]^{*}d_{J_{3}k_{J_{3}}}^{m}\left(x,y\right)d_{J_{4}k_{J_{4}}}^{n}\left(x,y\right)\right],
\label{eq:zeta3}
\end{align}
or
\begin{equation}
\zeta_{J_{1}J_{2}J_{3}J_{4}}^{\left(3\right)}\left(z\right)	=\frac{3}{\hbar}\sqrt{\frac{\left(\hbar\bar{\omega}_{J_{1}}\right)\left(\hbar\bar{\omega}_{J_{2}}\right)\left(\hbar\bar{\omega}_{J_{3}}\right)\left(\hbar\bar{\omega}_{J_{4}}\right)}{2^{4}\varepsilon_{0}^{2}\left[\mathcal{A}_{J_{1}J_{2}J_{3}J_{4}}^{\left(3\right)}\right]^{2}}}\frac{s\left(z\right)\bar{\chi}^{\left(3\right)}}{\bar{n}^{4}},
\end{equation}
where we have introduced
\begin{align}
\mathcal{A}_{J_{1}J_{2}J_{3}J_{4}}^{\left(3\right)}	&=\left|\int\text{d}x\text{d}y\frac{\chi_{jlmn}^{\left(3\right)}\left(x,y\right)\bar{n}^{4}\left[d_{J_{1}k_{J_{1}}}^{j}\left(x,y\right)\right]^{*}}{\varepsilon_{0}^{2}\bar{\chi}^{\left(3\right)}n^{2}\left(x,y;\bar{\omega}_{J_{1}}\right)n^{2}\left(x,y;\bar{\omega}_{J_{2}}\right)}\right.\nonumber\\
&\quad\times\left.\frac{\left[d_{J_{2}k_{J_{2}}}^{l}\left(x,y\right)\right]^{*}d_{J_{3}k_{J_{3}}}^{m}\left(x,y\right)d_{J_{4}k_{J_{4}}}^{n}\left(x,y\right)}{n^{2}\left(x,y;\bar{\omega}_{J_{3}}\right)n^{2}\left(x,y;\bar{\omega}_{J_{4}}\right)}\right|^{-1},
\end{align}
and put
\begin{align}\label{eq:chi3}
\Gamma_{jlmn}^{\left(3\right)}\left(\bm{r}\right)\rightarrow &\frac{\chi_{jlmn}^{\left(3\right)}\left(x,y\right)s\left(z\right)n^{-2}\left(x,y;\bar{\omega}_{J_{4}}\right)}{\varepsilon_{0}^{2}n^{2}\left(x,y;\bar{\omega}_{J_{1}}\right)n^{2}\left(x,y;\bar{\omega}_{J_{2}}\right)n^{2}\left(x,y;\bar{\omega}_{J_{3}}\right)},
\end{align}
ignoring the contribution of $\chi^{(2)}$ to $\Gamma^{(3)}$ \cite{quesada2019efficient}. Here $s\left(z\right)$ defines the nonlinear region of the waveguide, and $\bar{n}$ and $\bar{\chi}^{3}$ are, respectively, a refractive index and third-order susceptibility introduced solely for convenience. We point out that $\zeta_{JJJJ}^{\left(3\right)}\left(z\right)$ is connected to the more common nonlinear parameter~\cite{agrawal2007nonlinear} $\gamma$ with units of $\text{m}^{-1}\text{W}^{-1}$ via
\begin{equation}
\frac{\zeta_{PPPP}^{\left(3\right)}\left(z\right)}{s\left(z\right)}=\gamma\hbar\bar{\omega}_{P}v_{P}^{2}.
\end{equation}

The general operator equation resulting from Eqs.~\eqref{eq:HLpsi} and~\eqref{eq:HSPSFWM} of the main text is
\begin{align}
&\left(\frac{\partial}{\partial t}+v_{P}\frac{\partial}{\partial z}-i\frac{v_{P}^{\prime}}{2}\frac{\partial^{2}}{\partial z^{2}}\right)\bar{\psi}_{P}\left(z,t\right)\nonumber\\
&\qquad=i\zeta^{\left(3\right)}\left(z\right)\bar{\psi}_{P}^{\dagger}\left(z,t\right)\bar{\psi}_{P}\left(z,t\right)\bar{\psi}_{P}\left(z,t\right).
\end{align}
However, replacing $\bar{\psi}_{P}\left(z,t\right)$ by its mean field plus fluctuations $\bar{\psi}_{P}\left(z,t\right)\rightarrow\left\langle \bar{\psi}_{P}\left(z,t\right)\right\rangle +\delta\bar{\psi}_{P}\left(z,t\right)$, and neglecting terms quadratic in $\delta\bar{\psi}_{P}\left(z,t\right)$ leads to Eqs.~\eqref{eq:psiP} and~\eqref{eq:dpsiP} of the main text.

\subsection{Dual-pump SFWM}
Our third-order nonlinear coupling parameter $\zeta_{J_{1}J_{2}J_{3}J_{4}}^{\left(3\right)}\left(z\right)$ is given by Eq.~\eqref{eq:zeta3} as above. The general operator coupled-mode equations resulting from Eqs.~\eqref{eq:HLpsi} and~\eqref{eq:HDPSFWM} of the main text are
\begin{align}
&\left(\frac{\partial}{\partial t}+v_{P_{1}}\frac{\partial}{\partial z}-i\frac{v_{P_{1}}^{\prime}}{2}\frac{\partial^{2}}{\partial z^{2}}\right)\bar{\psi}_{P_{1}}\left(z,t\right)\nonumber\\	&\qquad=i\zeta_{P_{1}P_{1}P_{1}P_{1}}^{\left(3\right)}\left(z\right)\bar{\psi}_{P_{1}}^{\dagger}\left(z,t\right)\bar{\psi}_{P_{1}}\left(z,t\right)\bar{\psi}_{P_{1}}\left(z,t\right)\nonumber\\
&\qquad\quad+2i\zeta_{SP_{1}SP_{1}}^{\left(3\right)}\left(z\right)\bar{\psi}_{S}^{\dagger}\left(z,t\right)\bar{\psi}_{S}\left(z,t\right)\bar{\psi}_{P_{1}}\left(z,t\right)\nonumber\\
&\qquad\quad+2i\zeta_{P_{1}P_{2}P_{1}P_{2}}^{\left(3\right)}\left(z\right)\bar{\psi}_{P_{2}}^{\dagger}\left(z,t\right)\bar{\psi}_{P_{1}}\left(z,t\right)\bar{\psi}_{P_{2}}\left(z,t\right)\nonumber\\
&\qquad\quad+i\left[\zeta_{SSP_{1}P_{2}}^{\left(3\right)}\left(z\right)\right]^{*}\bar{\psi}_{P_{2}}^{\dagger}\left(z,t\right)\bar{\psi}_{S}\left(z,t\right)\bar{\psi}_{S}\left(z,t\right),  
\end{align}
\begin{align}
&\left(\frac{\partial}{\partial t}+v_{P_{2}}\frac{\partial}{\partial z}-i\frac{v_{P_{2}}^{\prime}}{2}\frac{\partial^{2}}{\partial z^{2}}\right)\bar{\psi}_{P_{2}}\left(z,t\right)\nonumber\\
&\qquad=i\zeta_{P_{2}P_{2}P_{2}P_{2}}^{\left(3\right)}\left(z\right)\bar{\psi}_{P_{2}}^{\dagger}\left(z,t\right)\bar{\psi}_{P_{2}}\left(z,t\right)\bar{\psi}_{P_{2}}\left(z,t\right)\nonumber\\
&\qquad\quad+2i\zeta_{SP_{2}SP_{2}}^{\left(3\right)}\left(z\right)\bar{\psi}_{S}^{\dagger}\left(z,t\right)\bar{\psi}_{S}\left(z,t\right)\bar{\psi}_{P_{2}}\left(z,t\right)\nonumber\\
&\qquad\quad+2i\zeta_{P_{1}P_{2}P_{1}P_{2}}^{\left(3\right)}\left(z\right)\bar{\psi}_{P_{1}}^{\dagger}\left(z,t\right)\bar{\psi}_{P_{1}}\left(z,t\right)\bar{\psi}_{P_{2}}\left(z,t\right)\nonumber\\
&\qquad\quad+i\left[\zeta_{SSP_{1}P_{2}}^{\left(3\right)}\left(z\right)\right]^{*}\bar{\psi}_{P_{1}}^{\dagger}\left(z,t\right)\bar{\psi}_{S}\left(z,t\right)\bar{\psi}_{S}\left(z,t\right),
\end{align}
and
\begin{align}
&\left(\frac{\partial}{\partial t}+v_{S}\frac{\partial}{\partial z}-i\frac{v_{S}^{\prime}}{2}\frac{\partial^{2}}{\partial z^{2}}\right)\bar{\psi}_{S}\left(z,t\right)\nonumber\\	&\qquad=i\zeta_{SSSS}^{\left(3\right)}\left(z\right)\bar{\psi}_{S}^{\dagger}\left(z,t\right)\bar{\psi}_{S}\left(z,t\right)\bar{\psi}_{S}\left(z,t\right)\nonumber\\
&\qquad\quad+2i\zeta_{SP_{1}SP_{1}}^{\left(3\right)}\left(z\right)\bar{\psi}_{P_{1}}^{\dagger}\left(z,t\right)\bar{\psi}_{S}\left(z,t\right)\bar{\psi}_{P_{1}}\left(z,t\right)\nonumber\\
&\qquad\quad+2i\zeta_{SP_{2}SP_{2}}^{\left(3\right)}\left(z\right)\bar{\psi}_{P_{2}}^{\dagger}\left(z,t\right)\bar{\psi}_{S}\left(z,t\right)\bar{\psi}_{P_{2}}\left(z,t\right)\nonumber\\
&\qquad\quad+2i\zeta_{SSP_{1}P_{2}}^{\left(3\right)}\left(z\right)\bar{\psi}_{S}^{\dagger}\left(z,t\right)\bar{\psi}_{P_{1}}\left(z,t\right)\bar{\psi}_{P_{2}}\left(z,t\right).
\end{align}
However, replacing $\bar{\psi}_{P_{1}}\left(z,t\right)$ and $\bar{\psi}_{P_{2}}\left(z,t\right)$ by their mean fields $\bar{\psi}_{P_{1}}\left(z,t\right)\rightarrow\left\langle\bar{\psi}_{P_{1}}\left(z,t\right)\right\rangle$ and $\bar{\psi}_{P_{2}}\left(z,t\right)\rightarrow\left\langle\bar{\psi}_{P_{2}}\left(z,t\right)\right\rangle$, and neglecting terms quadratic in $\bar{\psi}_{S}\left(z,t\right)$, leads to Eqs.~\eqref{eq:psiP1},~\eqref{eq:psiP2}, and~\eqref{eq:psiS} of the main text.

\section{Equations of motion for the moments} \label{sec:momentproc}
Starting from Eq.~\eqref{eq:FTQFields} of the main text and using the product rule one can easily find the equations of motion for the quadratics $b(\kappa,t)b(\kappa',t)$ and $b^\dagger(\kappa',t)b(\kappa,t)$. In the Heisenberg picture they are 
\begin{align}
\frac{\partial}{\partial t} b^\dagger(\kappa')b(\kappa) &= i\left(\omega(\kappa')-\omega(\kappa) \right)  b^\dagger(\kappa')b(\kappa) \\
&-i \int \frac{\dd{\kappa'''}}{\sqrt{2\pi}} \mathcal{S}^*(\kappa'+\kappa''') b(\kappa''') b(k)  \nonumber\\
&+i \int \frac{\dd{\kappa''}}{\sqrt{2\pi}} \mathcal{S}(\kappa+\kappa'') b^\dagger(\kappa'') b^\dagger(k) \nonumber\\
&-2 i  \int \frac{\dd{\kappa'''}}{\sqrt{2\pi}} \mathcal{M}^*(\kappa'-\kappa''') b^\dagger(\kappa''') b(k)  \nonumber\\
&+2 i  \int \frac{\dd{\kappa''}}{\sqrt{2\pi}} \mathcal{M}(\kappa'-\kappa'') b^\dagger(\kappa) b(k''),  \nonumber
\end{align}
and 
\begin{align} \label{eq:eomM}
\frac{\partial}{\partial t} b(\kappa)b(\kappa') &= -i \left(\omega(\kappa)+\omega(\kappa')\right) b(\kappa)b(\kappa') \\
&+i \int  \frac{\dd{\kappa'''}}{\sqrt{2\pi}} \mathcal{S}(\kappa'+\kappa''') b^\dagger(\kappa''')b(\kappa) \nonumber \\
&+i \int  \frac{\dd{\kappa''}}{\sqrt{2\pi}} \mathcal{S}(\kappa+\kappa'') b^\dagger(\kappa'')b(\kappa') \nonumber \\
&+2i \int  \frac{\dd{\kappa'''}}{\sqrt{2\pi}} \mathcal{M}(\kappa'-\kappa''') b(\kappa''')b(\kappa) \nonumber \\
&+2i \int  \frac{\dd{\kappa''}}{\sqrt{2\pi}} \mathcal{M}(\kappa-\kappa'') b(\kappa'')b(\kappa') \nonumber \\
&+  \frac{i}{\sqrt{2\pi}} \mathcal{S}(\kappa+\kappa'), \nonumber 
\end{align}
where, to simplify notation, we have not written the second argument (time) of the operators $b(\kappa,t)$ and $b^\dagger(\kappa,t)$ and the functions $\mathcal{S}(\kappa,t)$ and $\mathcal{M}(\kappa,t)$. Note that these equations are equally valid for the the moments $\braket{b(\kappa)b(\kappa')}$ and $\braket{b^\dagger(\kappa')b(\kappa)}$ which are precisely the expectation values of the quadratics above.
Note the last (inhomogenoeus) term in Eq.~\eqref{eq:eomM} appears due to normal ordering $b(\kappa) b^\dagger(\kappa''') = \delta(\kappa-\kappa''')+b^\dagger (\kappa''')b(\kappa)$.
This is the term that drives vacuum fluctuations into the system and that is responsible for generating squeezed vacuum if the input of the waveguide is vacuum.

At the moment level one can add loss by simply adding to the equations of motion derived before the rate of change of the moments due to loss. This is easily found from Eqs.~\eqref{eq:FieldsUpdateLoss} and \eqref{eq:expdecay} of the main text to be
\begin{align}
\left.\left( \frac{\partial}{\partial t} \braket{b^\dagger(\kappa') b(\kappa)} \right)\right|_{\text{loss}} &= -\gamma \braket{b^\dagger(\kappa') b(\kappa)},\\
\left.\left( \frac{\partial}{\partial t} \braket{b(\kappa) b(\kappa')} \right)\right|_{\text{loss}} &= -\gamma \braket{b(\kappa) b(\kappa')}.
\end{align}
We can put the unitary evolution due to the nonlinearity together with the rate of loss for the moment to write 
\begin{align}
\frac{\partial}{\partial t} \braket{b(\kappa)b(\kappa')} &= -i \left(\omega(\kappa)+\omega(\kappa') - i \gamma\right) \braket{b(\kappa)b(\kappa')} \nonumber \\
&+i \int  \frac{\dd{\kappa'''}}{\sqrt{2\pi}} \mathcal{S}(\kappa'+\kappa''') \braket{b^\dagger(\kappa''')b(\kappa)} \nonumber \\
&+i \int  \frac{\dd{\kappa''}}{\sqrt{2\pi}} \mathcal{S}(\kappa+\kappa'') \braket{b^\dagger(\kappa'')b(\kappa')} \nonumber \\
&+2i \int  \frac{\dd{\kappa'''}}{\sqrt{2\pi}} \mathcal{M}(\kappa'-\kappa''') \braket{b(\kappa''')b(\kappa)} \nonumber \\
&+2i \int  \frac{\dd{\kappa''}}{\sqrt{2\pi}} \mathcal{M}(\kappa-\kappa'') \braket{b^\dagger(\kappa'')b(\kappa')} \nonumber \\
&+  \frac{i}{\sqrt{2\pi}} \mathcal{S}(\kappa+\kappa').
\end{align}
Assuming that at some earlier time $t_0=-T/2$ one had $\braket{b(\kappa,t_0)b(\kappa',t_0)} = \braket{b^\dagger(\kappa,t_0)b(\kappa',t_0)} = 0$ one can solve the last equation perturbatively to a final time $t_1=T/2$
\begin{align}\label{eq:MandS}
\braket{b(\kappa,t_1)b(\kappa',t_1)} =  \frac{i}{\sqrt{2 \pi}} \int_{t_0}^{t_1}\dd{t}& \ e^{-i (\omega(\kappa)+\omega(\kappa')-i \gamma )(t_1-t)} \nonumber \\
&\times \mathcal{S}(\kappa+\kappa',t).
\end{align}
Now we specialize to the case of SPDC for which we can write the lossy-evolution of the pump field as in Eq.~\eqref{eq:pumpevol} to find
\begin{align}
\mathcal{S}(\kappa,t) =& \int \frac{\dd{z}}{\sqrt{2\pi}} e^{- i \kappa z} \zeta^{(2)}(z) \int \frac{\dd{\kappa''}}{\sqrt{2 \pi}} e^{i \kappa'' z} \times \\
&   e^{-i (\omega_{SH}(\kappa'') - i \gamma_{SH}/2)(t-t_0)} \braket{b_{SH}(\kappa'',t_0)}. \nonumber 
\end{align}
Using this last expression in Eq.~\eqref{eq:MandS} together with the definitions of the phase matching function in Eq.~\eqref{eq:pmf} and the complex energies in Eq.~\eqref{eq:cenergies} one arrives at the expression of Eq.~\eqref{eq:SPDCmoment} for the perturbative joint spectral amplitude in SPDC after performing a straightforward time integration.

\bibliographystyle{unsrt}
\bibliography{squeezing}

\end{document}